# High spatial resolution charge sensing of quantum Hall states

**Authors:** Cheng-Li Chiu[1], Taige Wang[2,3], Ruihua Fan[2], Kenji Watanabe[4], Takashi Taniguchi[5], Xiaomeng Liu[6], Michael P. Zaletel[2,3], Ali Yazdani[1]*

**Affiliations:**

[1] Joesph Henry Laboratories and Department of Physics, Princeton University, Princeton, NJ 08544, USA

[2] Department of Physics, University of California at Berkeley, Berkeley, CA 94720, USA

[3] Material Science Division, Lawrence Berkeley National Laboratory, Berkeley, CA 94720, USA

[4] Research Center for Functional Materials, National Institute for Materials Science, 1-1 Namiki, Tsukuba 305-0044, Japan

[5] International Center for Materials Nanoarchitectonics, National Institute for Materials Science, 1-1 Namiki, Tsukuba 305-0044, Japan

[6] Laboratory of Atomic and Solid State Physics, Cornell University, Ithaca, NY 14853, USA

*Correspondence to: yazdani@princeton.edu

**Charge distribution offers a unique fingerprint of important properties of electronic systems, including dielectric response, charge ordering and charge fractionalization. We develop a new architecture for charge sensing in two-dimensional electronic systems in a strong magnetic field. We probe local change of the chemical potential in a proximitized detector layer using scanning tunneling microscopy (STS), allowing us to infer the chemical potential and the charge profile in the sample. Our technique has both high energy (<0.3 meV) and spatial (<10 nm) resolution exceeding that of previous studies by an order of magnitude. We apply our technique to study the chemical potential of quantum Hall liquids in monolayer graphene under high magnetic fields and their responses to charge impurities. The chemical potential measurement provides a local probe of the thermodynamic gap of quantum Hall ferromagnets and fractional quantum Hall states. The screening charge profile reveals spatially oscillatory response of the quantum Hall liquids to charge impurities, and is consistent with the composite Fermi liquid picture close to the half-**

**filling. Our technique also paves the way to map moiré potentials, probe Wigner crystals, and investigate fractional charges in quantum Hall and Chern insulators.**

Multiple techniques have been established to spatially resolve electrostatic and chemical potential in two-dimensional electronic systems, including Kelvin probe force microscopy[1,2] (KPFM), electrostatics force microscopy[3,4] (EFM), scanning quantum dot microscopy[5] (SQDM), and the scanning single electron transistor[6,7] (SET). We have witnessed their recent success in understanding strongly correlated electronic systems, such as Wigner crystals[8], generalized Wigner crystal[9,10], topological phases in moiré heterostructures[11–13] and many more. However, high spatial resolution and high energy resolution have never been achieved simultaneously. For example, scanning SET, though offers remarkable energy resolution (<50 µeV), has limited spatial resolution (>100 nm) due to detector size. KPFM and SQDM can achieve a higher spatial resolution (<40 nm), but at the cost of limited energy resolution (>5 meV).

Facing these challenges, we develop a chemical potential and charge sensing technique with both high spatial and high energy resolution based on scanning tunneling spectroscopy (STS). We take advantage of the high spatial resolution of the conventional STS and convert it into a chemical potential probe by adding a detector layer whose spectroscopy reflects the chemical potential of the sample. Specifically, we choose the detector to be monolayer graphene (MLG), which has proven useful for imprinting the charge density variation of the sample layer[9,14,15]. Here, we fix the MLG to be a particular incompressible state such that its chemical potential is locked to that of the sample, which promotes the detector to a full-fledged quantitative chemical potential probe. The high spatial and energy resolution is achieved by preparing the detector MLG into an integer quantum Hall (IQH) state. The Landau levels (LL) can be used as a sharp spectroscopic feature for chemical potential readout, offering high energy resolution; the incompressible nature of the IQH state allows the detector to be put close to the sample while remaining non-invasive, making high spatial resolution possible. We use STS to track the energy of the LLs in the detector MLG, which can be converted to the chemical potential and charge distribution of the sample. Although using a detector layer in the IQH state requires a magnetic field, we found our technique viable at fields as low at 0.2T (see supplement section 9). Interestingly, while our experimental setup is similar to that of ref. 9, 10 and 14, the mechanisms for contrast differ: Ref. 9, 10 and 14 utilize local tip-gating effects to infer the charge profile of the sample layer from discharge events, whereas we rely on the band-bending effect in the detector layer to directly read out the chemical potential of the sample layer. Both effects are present in our setup, but we configure the STM tip and the electronic state of the detector layer such that the band-bending effect dominates, unlike ref. 9, 10 and 14, which operate in a regime where the local tip-gating effect is more prominent (see supplement section 10 for a full comparison).

**Device Structure and Control Parameters**

As shown in Fig. 1a, our device consists of two MLG layers: one serving as the detector and the other below as the sample, separated by a thin hexagonal Boron Nitride (hBN) layer (2.5 nm). A graphite back gate is 70 nm beneath the sample, separated by a thicker hBN layer. We fabricate our devices using state-of-the-art Van der Waals stacking techniques[16] and cleaning techniques which leave minimal residue on the exposed top layer MLG. The stack is placed onto a $SiO_2$/Si substrate with prepatterned electrical contacts. The filling of the detector and sample can be controlled individually by the gate voltages $V_g$ and $V_m$.

We use STM to measure the zero-bias tunneling conductance of the detector layer as a function of $V_g$ and $V_m$, in the presence of 6 T magnetic field at 1.4 K. As shown in Fig.1b, we can see compressible regimes (white) and incompressible quantum Hall regimes (red) of the detector layer. When we change the gate voltages, the sample and detector layers vary across different integer quantum Hall plateaus independently, giving rise to the zigzag pattern. These measurements match both the penetration capacitance measurements[17] (filling factor difference < 2%) of the same device and simulation with an electrostatic charging model (see supplement section 2 and 4). The excellent match highlights the non-invasive nature of our technique, and combined with the pristine quality of our device, enables the first-time STM observation of quantum Hall excitonic insulating states (for STS and penetration capacitance measurement of excitonic insulating states, see supplement section 3 and 4). \footnote{The 40mV shift in the double charge neutrality point (charge neutral point of both layers), indicated by a dot in Fig. 1b, arises from the built-

in electric field due to the device's asymmetry, where the top layer is single-side encapsulated by hBN.}

**Chemical potential sensing**

The basis of our chemical potential sensing technique is to measure the energy shift of the detector's LLs as we tune the back gate $V_g$ (see supplement section 6). Following the dashed line in Fig. 1b, we tune the detector MLG into an incompressible IQH state at $\nu = -2$ to minimize the tip-sample chemical potential mismatch[18] (see supplement section 5 for details), and charge the sample MLG to partial fillings of the zeroth Landau level (zLL). As shown in Fig 1c, the energy of the detector MLG's LL does not increase linearly with the back gate voltage as one would expect from a conventional MLG device[18,19]. In fact, the LL energy tracks the chemical potential $\mu_d$ of the detector MLG, which is locked to the chemical potential $\mu_m$ of the sample MLG. In a uniform sample, it can be shown from an electrostatic charging model that (see supplement section 6)

$$\mu_d = \mu_m + V_m - \frac{n_d}{c_d}$$

where $n_d$ is the charge density of the detector layer, and $c_d$ is the geometric capacitance $c_d = \epsilon/d_d$ per unit area between the detector and the sample. Both $V_m$ and $n_d$ remain the same when we charge the sample layer, so $\mu_d$ can be easily converted to $\mu_m$.

This locking behavior can also be understood from a multi-step process shown in Fig. 1d. When we apply a positive $\Delta V_g$ to the back gate (step 1) an electrochemical potential shift (step 2) will induce charge transfer between the back gate and the sample layer, increasing the sample layer's filling factor (step 3) and modifying its internal chemical potential $\mu_m$ (step 4). Meanwhile, the electrochemical potential difference between the detector and the sample layer is held constant by $V_m$ (Step 5). Since the detector is in an incompressible state, no charge transfer will occur between the sample and the detector. Consequently, the internal chemical potential of the detector layer will be pinned to that of the sample layer (step 6), tracing the internal chemical potential of the sample layer as we change the sample layer's filling factor. Since all the charge transfer occurs between the back gate and the sample layer, the change in filling factor will be proportional to $\Delta V_g$.

The resulting chemical potential $\mu_m$ of the sample MLG is presented in Fig. 1e as a function of filling factor. The compressibility is negative within each LL due to electron-electron interactions[17,20,21]. We have also observed charge gaps associated with the symmetry-breaking quantum Hall ferromagnetism (QHFM) in the zLL and the fractional quantum Hall (FQH) states (at $\nu = -\frac{1}{3}, -\frac{2}{3}, -\frac{2}{5}, -\frac{3}{5}$) within each isospin flavor of the zLL (see Fig. 1f for a closer look at FQH in both $N = 0$ and $N = 1$ LL). The QHFM and FQH gaps, measured by the jumps in the chemical potential, are comparable with previous measurements using capacitance techniques[15,22] and STM[18].

**Charge Sensing and spatially resolved response to single impurities**

Now we turn our attention to the proximity of a local impurity. The electrostatic charging model suggests that local changes in the LL energy $\Delta\mu_d(r)$ can be viewed as a probe of local changes in electrostatic potential $\Delta\Phi_d(r)$ (see supplement section 7),

$$\Delta\mu_d(r) = -\Delta\Phi_d(r)$$

where $\Delta\mu_d(r) = \mu_d(r) - \mu_d^0$, and $\mu_d^0$ is the chemical potential far away from the impurity (see Eq. 1). Physically, $\Delta\Phi_d(r)$ directly characterizes the screening property of the sample MLG. It has two contributions, one from the impurity itself $\Phi_d^{imp}(r)$ and one from the sample MLG's response $\Phi_d^m(r)$, only the latter of which contains information about the sample. We show a two-dimensional map of $\Delta\Phi_d(r)$ at $\nu = -\frac{1}{2}$ in the inset of Fig. 2a (see methods). The positive charge impurities show up as dark disks in $\Delta\Phi_d(r)$, lowering the electrostatic potential for electrons. \footnote{We find the density of such impurities in our sample to be around 3x10⁹ cm⁻², which is comparable to that of the double-encapsulated devices.}

A closer look at an individual charge impurity (Fig. 2a, impurity A) reveals oscillations of the underscreening (dark) and overscreening (bright) regions. The oscillatory charge response in a partially filled LL can have multiple origins, including charge order instabilities like melted Wigner crystals[8], magnetoroton excitations[23], or Friedel oscillations of the composite Fermi liquid[24], depending on the filling factor. At $\nu = -\frac{1}{2}$, we show the Fourier transform of $\Delta\Phi_d(r)$ in Fig. 2f, which reveals a peak at wave vector close to $2/\ell_B$. The inset of Fig. 2c shows a scaling collapse of $\Delta\Phi_d(r)$ at various magnetic field by a scaling function $F(x) = \Delta\Phi_d(x\ell_B)/E_c$, where the magnetic length $\ell_B$ is the only length scale and the Coulomb energy $E_c = \frac{e^2}{\epsilon\ell_B}$ is the only energy scale. Our finding is consistent with the composite fermion liquid (CFL) theory[24–30] of the half-filled LL, which predicts a peak at $2k_F$ in the charge response function, with $k_F = 1/\ell_B$ being the composite fermion Fermi wavevector assuming full spin and valley polarization. We further compare our experimental results to the random phase approximation (RPA) calculation based on CFL theory (Fig. 2g; see details in the supplement 14, 15) and exact diagonalization (ED) calculations (see supplement section 13). The experimental results agree qualitatively with the RPA calculation, both showing a peak at wavevector slightly smaller than $2/\ell_B$ due to finite temperature effect. The ED calculation shows that the oscillatory behavior of the charge response survives even with a very strong impurity like impurity A.

To extract the actual charge response of the sample MLG, we need to isolate the electrostatic potential $\Phi_d^m(r)$ from the screening charges in the sample MLG. To do so, we first tune the sample MLG to the incompressible $\nu = -2$ IQH as well so that it cannot have any response $\Phi_d^m(r) = 0$. \footnote{More precisely, the sample MLG will exhibit only a dielectric response with 0 total charge, which we implicitly account for as a renormalization of $\Phi_d^{imp}(r)$.} Then we can measure $\Phi_d^{imp}(r)$ from $\Delta\Phi_d(r)$ directly. In this case, the spectroscopy at the impurity (Fig. 2b left) shows a set of discretized energy levels corresponding to different angular momentum states in the zLL of the detector. We can then estimate the depth of the impurity by matching these energy levels to numerical simulation. Assuming the impurity has unit charge +e, we find that the impurity is directly above the sample layer, probably introduced during the stacking process (Fig. 2b right, and see supplement section 11 for details). We can then isolate the potential $\Phi_d^m(r)$ arising from the sample MLG's response at arbitrary filling by subtracting $\Phi_d^{imp}(r)$ from $\Delta\Phi_d(r)$ (yellow line in Fig. 2b, and 2d). It becomes clear that the first peak in $\Delta\Phi_d(r)$ closest to impurity is associated with an overscreening effect as the screening charge is forced to spread out over $\ell_B$ and therefore $-\Phi_d^m(r)$ exceeds $\Phi_d^{imp}(r)$ at scale $\ell_B$ (see Fig. 2c). Using the electrostatic Green's function between sample and detector layers, we can further convert

$\Phi_d^m(r)$ into the charge distribution in the sample layer as shown in Fig. 2e. The screening charge density sums up to $-0.97 \pm 0.2 e$ (see supplement section 12 for details), and decays from the center of the impurity with oscillations over 60 nm. From this analysis, we estimate that our technique has a sensitivity of 0.3 meV to changes in the potential landscape (at 1.4 K) with a spatial resolution of about one $\ell_B$ (see supplement section 8).

**Filling dependence and impurity dependence**

The screening property of the sample MLG exhibits a rich dependence on the filling factor and the impurity strength. Fig. 3a shows the result for the impurity A across the range $\nu \in (-1,0)$ as a concrete example. The total charge density is locally bounded by $1/2\pi\ell_B^2$ in a single LL, which constrains the mobility of the screening charge density at higher filling. Consequently, the central underscreening region increases monotonically in diameter with the filling factor. Notably, the oscillation is much weaker at rational fillings $\nu = -\frac{2}{3}, -\frac{1}{3}$, as expected the incompressible nature of FQH states. The azimutal anisotropy at these fillings might be related to an anisotropic arrangement of fractionalized quasi-particles at the impurity. The amplitude of the oscillation is relative weak close to $\nu = -\frac{1}{2}$, yet survives farthest away from the impurity and shows more oscillation periods. We therefore interpret the oscillations in the charge response close to $\nu = -\frac{1}{2}$ as a universal feature of underlying electronic system.

The behaviors of the oscillation period fall into two distinct categories depending on the strength of the impurity. For a strong impurity, e.g. the impurity A and B (shown in Fig 3b, 4a respectively), the oscillation period becomes shorter with increasing filling factor for $\nu < -\frac{2}{3}$ while longer for $\nu > -\frac{2}{3}$, which is clearly not particle-hole symmetric about $\nu = -\frac{1}{2}$. In contrast, a weak impurity (C), which is 4.5 nm below the sample, exhibits a more symmetric behavior about $\nu = -\frac{1}{2}$ (Fig 4b). Fourier transform of the data for $-0.6 < \nu < -0.4$ also shows distinctive behaviors: the strong impurity B features a peak at a wave vector which decreases with filling (Fig 4c), while the weak impurity C features a dip at almost constant wave vector $2/\ell_B$ (Fig 4d). These observations are consistent with the exact diagonalization calculation shown in Fig. 4e,f. Particle-hole symmetry of the predicted CFL at $\nu = -\frac{1}{2}$ is an important question both experimentally and theoretically[31–34]. Our results suggest that the particle-hole symmetry, unless explicitly broken by a strong impurity, is preserved in the zLL of MLG close to $\nu = -\frac{1}{2}$. Nevertheless, inferring particle-hole symmetry breaking from $2k_F$ scattering should be with caution since the residual magnetic field composite fermions experience away from $\nu = -\frac{1}{2}$ can smear out the $2k_F$ peak in the charge response[25].

In conclusion, we have developed a novel technique to detect the collective density response of quantum hall liquid that achieves high spatial and energy resolution simultaneously. Our technique can also be applied beyond quantum Hall effect in graphene and even be adapted to zero magnetic field if the sensor is replaced with an isolated flat band material, such as twisted-bilayer graphene[35,36]. This opens up a new way to probe more exotic two-dimensional materials, including detecting fractional excitation in fractional Chern insulator[11,37,38] (FCI), Hall crystals in pentalayer graphene[39,40] and so on.

## Methods:

### Sample preparation:

The sample in this work was fabricated using a mechanical dry transfer technique. Our pickup stamp is made of polyvinyl alcohol (PVA) coated transparent tape. We use the stamp to pick up the layer, beginning from the top layer (detector). After picking up all the flakes, we align and press the stack on PVA against Au/Ti patterned Si/SiO2 chips. After heating up to 110°C, we are able to detach the PVA from the tape, leaving the sample covered underneath. Afterward, we wash away the PVA to expose the sample with water, acetone, IPA, and N-methyl-2-pyrrolidone (NMP). The sample is then loaded into the STM's UHV chamber and baked at 350°C overnight before being loaded into the microscope of our STM. The optical image of our device is in the supplementary material.

### STM measurements:

In this study, we use a home-built UHV STM with a maximum 6T out-of-plane magnetic field operating at T=1.4K to perform multiple types of measurements, including point spectroscopy measurement, grid measurement (2D potential mapping), line cut measurement, and penetration capacitance measurement. The tip is made of tungsten wire and prepared on a Cu(111) surface as described in our previous work[17]. When performing the point spectroscopy measurement (fig. 1c), we set our lock-in frequency to 4 kHz, using 1.5 mV oscillation, and the setpoint is 1 nA at Bias=+0.5V. When performing the grid measurement (figs. 2a, 3a) and line cut (fig. 2b, 4), we set the lock-in frequency to 2022 Hz and read the dI/dV and $d^2I/dV^2$ using two synchronized lock-ins, with the setpoint being 1-1.5 nA at Bias=+0.5V. Details on using $d^2I/dV^2$ to perform 2D potential mapping are in the supplementary material section 7. When doing the penetration capacitance measurement, we use 27.186 kHz and 2 mV oscillation for inter-layer compressibility and 18.648 kHz and 4 mV oscillation for total compressibility. The results of the penetration capacitance measurements are in the supplementary material section 4.

## Acknowledgements

## Funding

This work was supported by the DOE-BES grant DE-FG02-07ER46419, EPiQS initiative grant GBMF 9469 of the Gordon and Betty Moore Foundation, NSF through the Princeton

University (PCCM) Materials Research Science and Engineering Center (MRSEC) DMR-2011750, NSF DMR-2312311, ARO MURI 134396-5117989 (AROW911NF2120147), ONR N00014-21-1-2592, and NSF OMA-2326767. T.W. and M.Z. are supported by the U.S. Department of Energy, Office of Science, Office of Basic Energy Sciences, Materials Sciences and Engineering Division under Contract No. DE-AC02-05-CH11231 (Theory of Materials program KC2301). T.W. is also supported by the Heising-Simons Foundation, the Simons Foundation, and NSF grant No. PHY-2309135 to the Kavli Institute for Theoretical Physics (KITP). R.F. is supported by the Gordon and Betty Moore Foundation (Grant GBMF8688). This research used the Lawrencium computational cluster provided by the Lawrence Berkeley National Laboratory (supported by the U.S. Department of Energy, Office of Basic Energy Sciences under Contract No. DE-AC02-05-CH11231).

**Competing interests:** Authors declare no competing interests.

**Data and materials availability**: The data that support the plots within this paper are available at the Harvard Dataverse. Other data that support the findings of this study are available from the corresponding authors upon reasonable request.

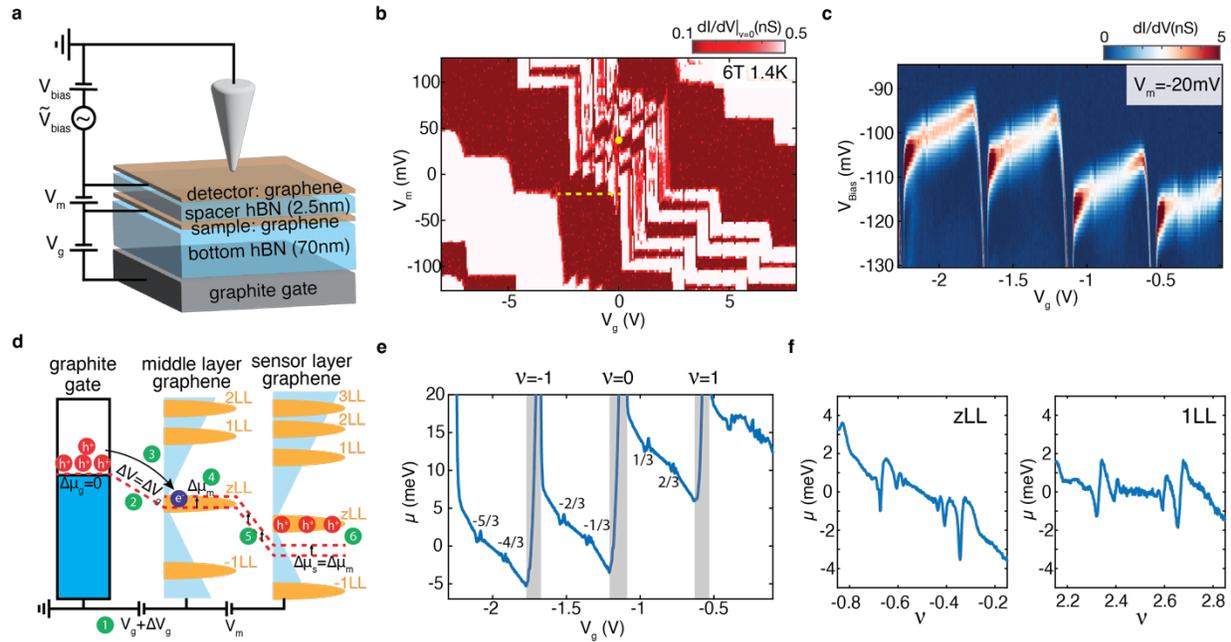

**Fig.1 Charge sensing technique and local chemical potential probing.** (a) Schematic of charge sensing measurement. $V_{bias}$ (DC and AC) is applied to the detector layer to control the tunneling while $V_m$ and $V_g$ are applied to the sample layer and the bottom gate to control the layer filling factor. (b) Zero bias conductance phase diagram showing the compressible and incompressible regimes. STM tip height is set at I=2.2nA, $V_{bias}$=0.4V before turning off the feedback and measure the zero bias conductance with $V_{ac}$=2mV. (c) Density-tuned bias spectroscopy of the -1LL. The Landau level traces the chemical potential of the sample layer. (d) Illustration of the charge sensing process. (e) Chemical potential extracted from (c) showing symmetry-breaking gaps and fractional quantum Hall gaps. (f) High-resolution measurements of the chemical potential of partially filled zLL and 1LL show many fractional quantum Hall states.

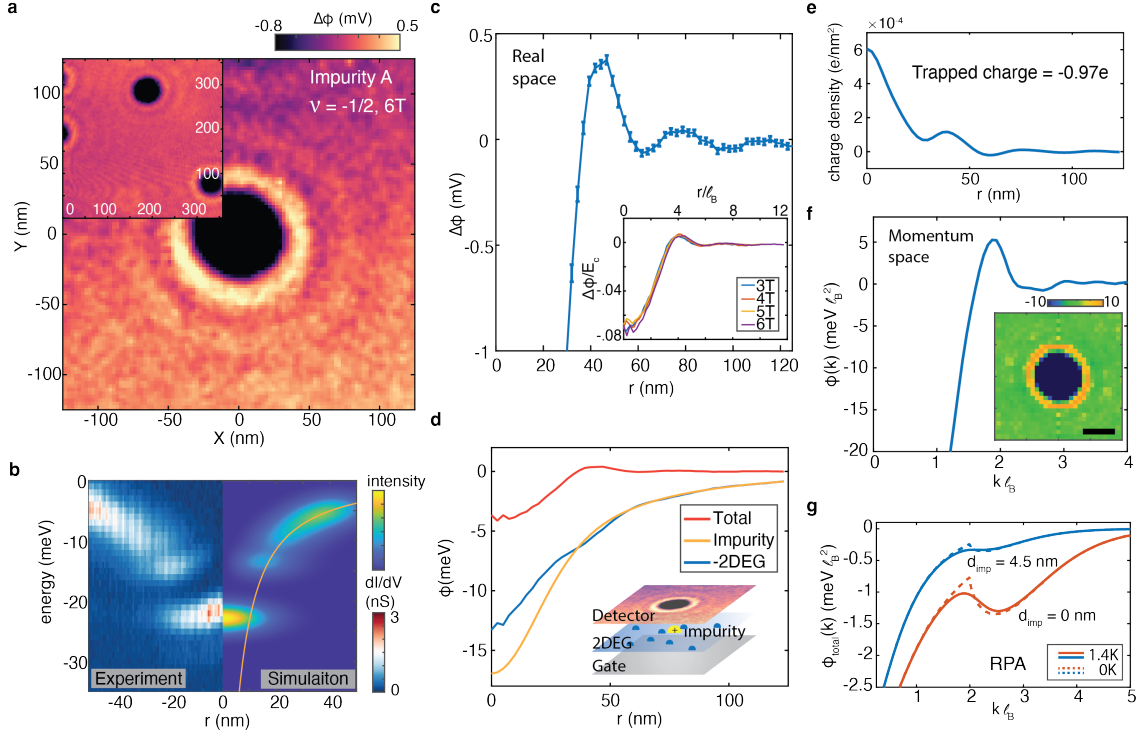

**Fig. 2 Spatial resolved potential response to a charge impurity at $v = -1/2$.** (a) Potential near a positive charge impurity taken at $v = -1/2$ under 6T. Inset: a wide field of view potential including multiple charge impurities taken with the same filling and magnetic field. (b) Left: Degeneracy-lifted state of zLL taken near the impurity when the sample and detector are at $v = +2$. The zero energy is shifted to the energy of the Landau level at a far distance. Right: Simulation using first-order perturbation theory with zLL wavefunction. The spectrum is broadened by 1.5 meV for clarity. (c) Azimuthal average of (a). The error bar represents the standard error of mean. Inset: Azimuthally averaged potential taken at different magnetic fields. The data is scaled with $E_c$ and $\ell_B$. (d) Potential profile of the impurity, 2DEG, and total. The screening potential from 2DEG is inverted for clarity. Inset: Illustration showing that the potential on the detector is the sum of impurity potential and screening potential. (e) Calculated charge density of the 2DEG screening the impurity potential. (f) Fourier transform of the potential. The result is azimuthally averaged from the 2D Fourier transform of (a). Inset: 2D Fourier transform of (a). (g) The composite fermion RPA calculation of the total potential at $v = -1/2$ for strong and weak potential. The dashed line and solid line represent the zero temperature and finite temperature (1.4K) calculations.

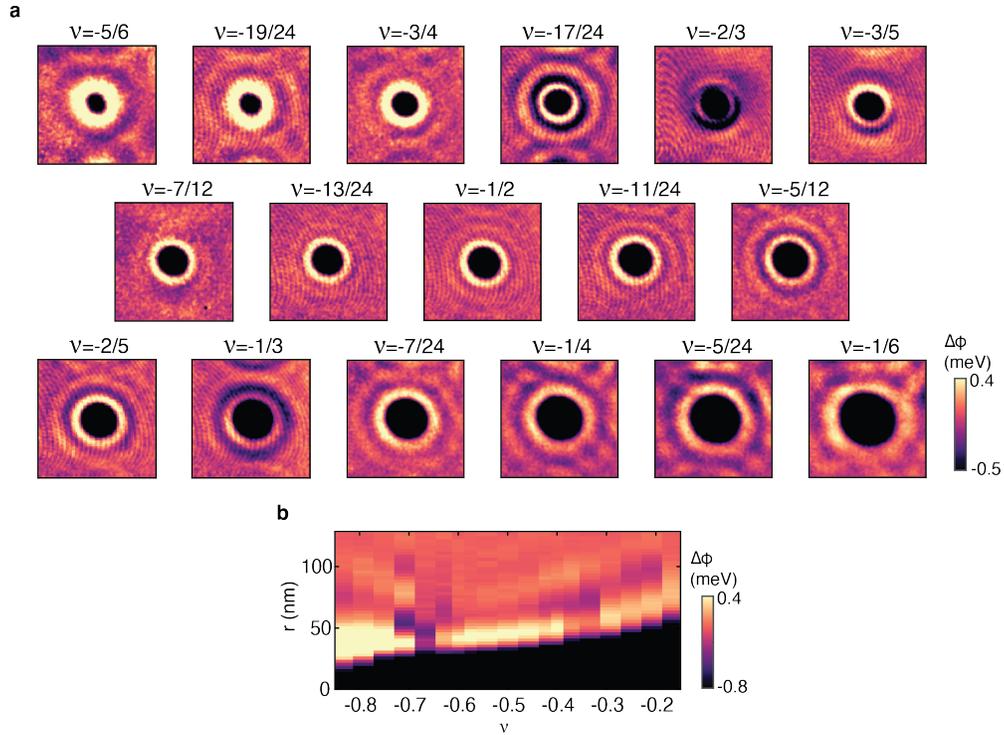

**Fig. 3 Potential response to a charge impurity at various filling factors.** (a) Potential taken at different filling factors. Each image is 252 x 252 nm with 100 x 100 pixels. Overscreening regions appear as bright color rings. The first row is $v = -5/6 \sim -3/5$. The second row is near half filling $v = -7/12 \sim -5/12$. The third row is $v = -2/5 \sim -1/6$. (b) Azimuthal average of potential as a function of filling factor. The dispersion of the potential shows the change of ring size and periodicity, which evolves with respect to $v$. The separation of the bright stripe, representing the periodicity of the oscillation ring, decreases before $v = -2/3$ and increase after $v = -3/5$.

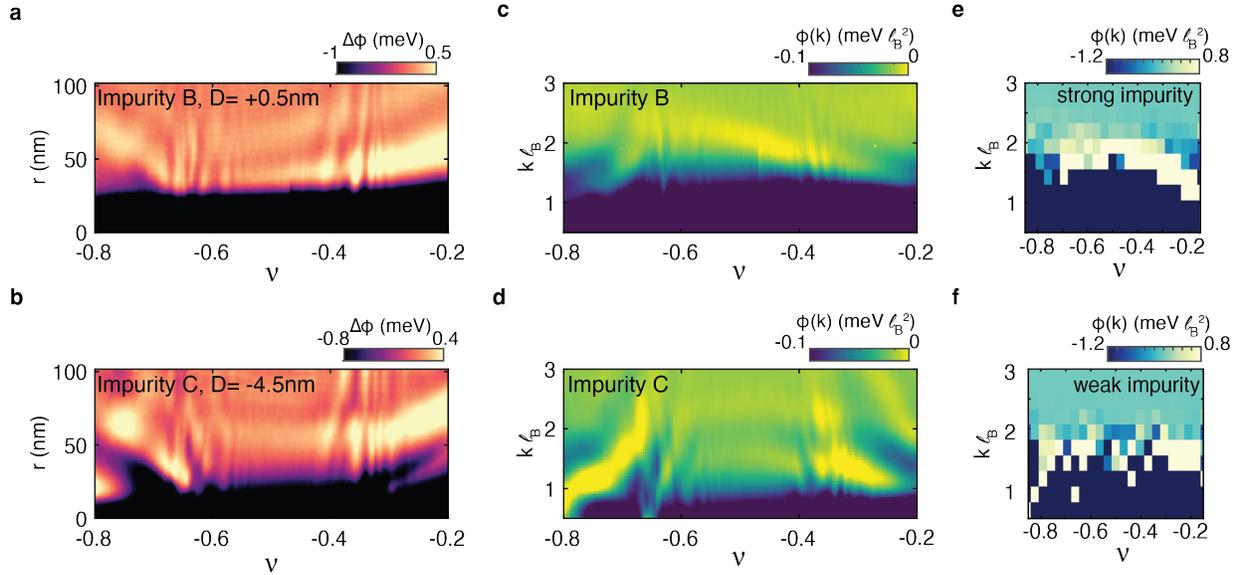

**Fig. 4 Potential response to different depths of charge impurities.** (a) Total potential near impurity B (2 nm below the detector, or 0.5 nm above the sample). The response shows similarity to impurity A. (b) Total potential near impurity C (7 nm below the detector, or 4.5 nm below the sample). The impurity potential is shallower since the impurity is further from the sample layer. This causes the response to be weaker compared to impurities A and B. (c) The Fourier transform of the total potential in (a). The response is asymmetric about $\nu = -1/2$. (d) The Fourier transform of the total potential in (b). The response is symmetric across $\nu = -1/2$. (e) ED calculation of $\Phi_d^m(k)$ responding to a strong potential. The peak response disperses asymmetrically about $\nu = -1/2$ similar to (c). (f) ED calculation of $\Phi_d^m(k)$ responding to a weak impurity exhibits a more symmetric response about $\nu = -1/2$, consistent with the experimental observation in (d). The apparent "noise" is an artifact coming from the interplay of the Jain sequence and the discrete sequence of $\nu$ sampled by finite-site numerics ($N_\phi = 30$) at T = 0.